\pdfoutput=1
\RequirePackage{ifpdf}
\ifpdf 
\documentclass[pdftex]{sigma}
\else
\documentclass{sigma}
\fi

\usepackage[all]{xy}

\numberwithin{equation}{section}

{\theoremstyle{definition}
}

\newcommand{\D}{\mathrm{d}}
\newcommand{\dsl}[1]{{\displaystyle{#1}}}
\newcommand{\RR}{{{\mathbb{R}}}}

\begin{document}

\newcommand{\arXivNumber}{1907.10920}

\renewcommand{\PaperNumber}{087}

\FirstPageHeading

\ShortArticleName{On the Geometry of Extended Self-Similar Solutions of the Airy Shallow Water Equations}

\ArticleName{On the Geometry of Extended Self-Similar Solutions\\ of the Airy Shallow Water Equations}

\Author{Roberto CAMASSA~$^\dag$, Gregorio FALQUI~$^\ddag$, Giovanni ORTENZI~$^\ddag$ and Marco PEDRONI~$^\S$}

\AuthorNameForHeading{R.~Camassa, G.~Falqui, G.~Ortenzi and M.~Pedroni}

\Address{$^\dag$~University of North Carolina at Chapel Hill, Carolina Center for Interdisciplinary Applied\\
\hphantom{$^\dag$}~Mathematics, Department of Mathematics, Chapel Hill, NC 27599, USA}
\EmailD{\href{mailto:camassa@amath.unc.edu}{camassa@amath.unc.edu}}

\Address{$^\ddag$~Dipartimento di Matematica e Applicazioni, Universit\`a di Milano-Bicocca, Milano, Italy}
\EmailD{\href{mailto:gregorio.falqui@unimib.it}{gregorio.falqui@unimib.it} \href{giovanni.ortenzi@unimib.it}{giovanni.ortenzi@unimib.it}} 

\Address{$^\S$~Dipartimento di Ingegneria Gestionale, dell'Informazione e della Produzione, \\
\hphantom{$^\S$}~Universit\`a di Bergamo, Dalmine (BG), Italy}
\EmailD{\href{mailto:marco.pedroni@unibg.it}{marco.pedroni@unibg.it}}

\ArticleDates{Received July 17, 2019, in final form October 31, 2019; Published online November 09, 2019}

\Abstract{Self-similar solutions of the so called Airy equations, equivalent to the dispersionless nonlinear Schr\"odinger equation written in Madelung coordinates, are found and studied from the point of view of complete integrability and of their role in the recurrence relation from a bi-Hamiltonian structure for the equations. This class of solutions reduces the PDEs to a finite ODE system which admits several conserved quantities, which allow to construct explicit solutions by quadratures and provide the bi-Hamiltonian formulation for the reduced ODEs.}


\Classification{37K05; 37J15; 76M55}

\section{Introduction}
Integrable PDEs in one spatial dimension are well known to admit reductions to invariant finite-dimensional manifolds, where they give rise to integrable (and sometimes solvable) systems of ODEs. Among the prototype examples of these reductions are the finite-dimensional manifolds of the stationary points of one of the vector fields of the integrable hierarchy of PDEs (see, e.g.,~\cite{AS81} and the references quoted therein). In this case, the restriction of the PDE to the invariant submanifold is a Hamiltonian system of ODEs. Bogoyavlenskii and Novikov~\cite{BN70} described a~far-reaching scheme to construct the Hamiltonian function of the reduced system in terms of the Hamiltonian of the original PDE by means of techniques coming from the classical calculus of variations. In general, a vast amount of literature is devoted to the study of these and related aspects of reductions to finite dimensional systems, see, e.g., \cite{ACFHM, CalDeg, RauchW} and references quoted therein.

In this work we discuss a specific instance, inspired by the problem of finding (generalized or, in the terminology of \cite{Sed}, of the {\em second kind}\/) self-similar solutions of the well known system of dispersionless PDEs
\begin{gather}
\eta_t+(\eta u)_x=0, \qquad u_t+uu_x+\eta_x=0.\label{SW0}
\end{gather}
This system of PDEs is one of the oldest and most important non-linear models for shallow water waves in incompressible fluid dynamics. It is referred to in the literature with many different names, such as Saint-Venant system~\cite{SaVe}, one-dimensional Benney system (see, e.g.,~\cite{Zak}), Stokes system~\cite{Stokes}, or shallow water equations {\em tout court}. We will henceforth refer to system~(\ref{SW0}) as the {\em Airy} shallow water equations, or simply Airy system, following \cite[Section~182]{Lamb} and Stoker's book \cite[Introduction]{Sto}, where the study of~(\ref{SW0}) in the seminal Airy's paper \cite{Airy} is pointed out.

Throughout the present paper, the function $\eta=\eta(x,t)$ describes the ``water'' layer thickness, and $u=u(x,t)$ is the layer-averaged horizontal velocity. As well known, this system can also be viewed as a model for polytropic gas dynamics, {for the special specific heat ratio $c_p/c_v=2$, so that the pressure is a quadratic function of the density (see, e.g., \cite{Whi})}. As we recall below, these equations admit the classical self-similar solutions of the dam-break or piston problem \cite{Sto,Whi}, whose simplest case is
\begin{gather}
\eta = \frac{1}{9} \left(\dsl{\frac{x}{t}}\right)^2 ,\qquad u = \frac{2}{3} \left(\dsl{\frac{x}{t}}\right) .\label{Lax-raref-intro}
\end{gather}
However, these classical rarefaction waves are not quite adequate to characterize interface motions in presence of dry or vacuum points $x=x_v$ say, where $\eta(x_v,t)=0$ and system~(\ref{SW0}) ceases to be purely hyperbolic (see \cite{CFOPT} for more details). In the spirit of seeking self-similar solutions of the second kind,
the same spatial scaling as in equations~(\ref{Lax-raref-intro}) can be maintained with different time scalings, which gives rise to a three-parameter family (already hinted at in~\cite{Ovs}). These self-similar solutions were shown to play a fundamental role in the study of classes of initial data with dry points in~\cite{CFOPP, CFOPT}.
The time scaling factor is assumed to be a non trivial function $\tau(t)$, leading to explicit solutions of the form
\begin{gather}\label{ovos}
\eta(x,t)=\gamma(t) x^2 +\zeta(t) , \qquad u(x,t)=\alpha(t) x ,
\end{gather}
where
\begin{gather}\alpha^2= 4 \gamma_0 \tau^3 + \big(\alpha_0^2 -4 \gamma_0\big) \tau^2 ,\qquad \gamma=\gamma_0 \tau^3 , \qquad \zeta=\zeta_0 \tau ,\label{gensol3cpi}
\end{gather}
and $\tau(t)$ solves
\begin{gather*}
\dot{\tau}^2= 4 \gamma_0 \tau^5 + \big(\alpha_0^2 -4 \gamma_0\big) \tau^4 , \qquad \tau(0)=1
\end{gather*}
(as we shall see in Section~\ref{sec:explicit-solutions}, replacing the independent variable $t$ with $\tau$ has an interesting geometrical interpretation).
Being nonlinear, these equations can have finite time singularities, which for the PDE~(\ref{SW0}) can be interpreted as the formation of gradient catastrophes and shock waves~\cite{CFOPP, CFOPT}. Solutions~(\ref{gensol3cpi}) are obtained by the study of conserved quantities. Their structure is deeply related to the bi-Hamiltonian recursion of the parent system~(\ref{SW0}). A thorough study of this connection from the geometrical viewpoint is one of the aims of this work.

A natural generalisation of the field configurations above (see \cite{Ovs}) describes a general para\-bo\-lic shape for the water height function $\eta(x,t)$ and a linear profile for the speed $u(x,t)$, and reads
\begin{gather}\label{ovo}
\eta(x,t)=\gamma(t) x^2+\omega(t) x +\zeta(t) , \qquad u(x,t)=\alpha(t) x+\beta(t) .
\end{gather}
As shown below, the simplicity of these relations allows the construction of explicit solutions and a thorough description of their geometrical properties. (The degenerate case of $\gamma\equiv0$ requires a separate study, as it is a singular limit of the reduction.) Further interest in such a reduction lies in the fact that the manifold~(\ref{ovo}) is in some sense an attractor under the action of the special Lie symmetry of (\ref{SW0}) given by
 \begin{gather*}
 \eta_s=2 \eta-{x\eta_x},\qquad u_s=u-xu_x .
\end{gather*}
More precisely, generalized self-similar solutions, which according to~\cite{BZ} describe intermediate asymptotic behaviour(s) of general solutions of PDEs, can be realized within this invariant manifold as suitable paths in the parameter space $(t,s)$, as discussed in Section~\ref{subsection:Lie symmetry}.

The organization of this work is as follows. In Section \ref{Scaling} we first discuss and recall the relevance of self-similar solutions of the Airy system, and recall some notions of the bi-Hamiltonian structure of the system in Section~\ref{bi-sub}. Next, we study the resulting ODEs in their own right. More precisely, in Section \ref{link} we find suitable conserved quantities, obtained by a well known sequence of global constants of the motion for the Airy system, and we then use them to explicitly solve, in Section~\ref{sec:explicit-solutions}, the ODEs system. In Section~\ref{sezbih} we find a bi-Hamiltonian structure of the reduced ODEs, by requiring that the above mentioned quantities give rise to a finite Lenard--Magri sequence. Finally, in Section~\ref{redux} we briefly discuss the reductions of the general ODE system to the space of parity preserving parabolic configurations, and to the degenerate space of linear-linear configurations.

\section[Self-similarity, their Lie symmetry and a bi-Hamiltonian setting for the Airy system]{Self-similarity, their Lie symmetry\\ and a bi-Hamiltonian setting for the Airy system}\label{Scaling}

Self-similarity is well known to provide interesting solutions to physically relevant PDEs (see, e.g., \cite[Chapter~IV]{Sed} and \cite{BZ}). First kind self-similar solutions of the Airy-SWE
 \begin{gather}
\eta_t+(\eta u)_x=0 , \qquad u_t+uu_x+\eta_x=0\label{SW1}
\end{gather}
are obtained via the following ansatz, based on dimensional analysis:
\begin{gather}\label{classAns}
\eta \left( x,t \right) =N(\xi) , \qquad u\left( x,t \right) =U(\xi) , \qquad \text{with}\quad \xi=\frac{x}{t} .
\end{gather}
By substituting this in (\ref{SW1}) and requiring matching of the time-exponents, we obtain the well known self-similar solutions of the piston problem \cite{Sto,Whi}
\begin{gather}
\eta = \frac{1}{9} (\xi-2C)^2 ,\qquad u = \frac{2}{3} (\xi+ C) , \qquad C \in \mathbb{R} .\label{Lax-raref}
\end{gather}
It is a classical result \cite{Sto,Whi} that self-similar solutions of the form (\ref{Lax-raref}) arise in the context of the Riemann (or dam-break) problem.
However, in \cite{LS} and \cite{GK} it is remarked that such behaviour (and its shock-like counterpart) seems to be inadequate to study analytic solutions near their critical points (in particular, the ``vacuum'' or ``dry'' points). Indeed, as pointed out by Barenblatt and Zel'dovich \cite{BZ}, sometimes dimensional analysis is not sufficient to find the right metastable evolutions in certain regimes, and a generalised notion of self-similarity (referred to as of the second kind) needs to be introduced.
This is defined through a solution of the form
\begin{gather}
u=A(\tau) U(\xi) , \qquad \eta=B(\tau) N(\xi) ,\label{gss}
\end{gather}
where
\begin{gather*}
\tau=\tau(t), \qquad \xi=x \tau,\qquad \mathrm{and} \qquad \frac{\D {\tau}}{\D t}=F(\tau) .
\end{gather*}
The functions $A(\tau)$, $B(\tau)$, $U(\xi)$, $N(\xi)$ and the time scaling $F(\tau)$ are unknown functions of one variable, to be determined by the equations of motion (\ref{SW1}), with the sole requirement that $F(\tau)$ be never vanishing. Upon the substitution~(\ref{gss}), one obtains
\begin{gather}
(B_\tau F) N + \frac{BF}{\tau} \xi N_\xi +(AB\tau) (UN)_\xi=0 , \nonumber\\
(A_\tau F) U + \frac{AF}{\tau} \xi U_\xi +\big(A^2\tau\big) UU_\xi+(B \tau) N_\xi=0 .\label{SW-ss}
\end{gather}
We notice that the mathematical structure of the resulting equations consists of the sums of terms that are factorized into a function of $\tau$ and a function of the scaling variable $\xi$,
\begin{gather}
\sum_i \Phi_i(\tau) \Psi_i(\xi)=0 .\label{structsep}
\end{gather}
The classical ansatz (\ref{classAns}) is equivalent to the requirement that in~(\ref{structsep}) all the functions $\Phi_i(\tau)$ coincide.
However, different choices can be made for the ``time'' factors $\Phi_i(\tau)$ in~(\ref{structsep}), leading to a consistent set of equations.
Let us still require proportionality in the first of~(\ref{SW-ss}), but choose two different $\tau$-scaling factors in the second one, as follows:
\begin{gather}\label{tau-2}
B_\tau F = k \frac{BF}{\tau}=l AB\tau , \qquad A_\tau F =m B \tau , \qquad \frac{A F}{\tau} = n A^2 \tau,
\end{gather}
where $k$, $l$, $m$, $n$ are real constants. This yields
\begin{gather*}
B= \tau_0 \tau^k , \qquad l=nk ,\qquad F=nA\tau^2 , \qquad A^2 =\frac{2m\tau_0}{nk}\tau^k+a_0 ,
\end{gather*}
for some constants $\tau_0$, $a_0$ (the latter is assumed non vanishing). Equations~(\ref{SW-ss}) now become an overdetermined system of {\em three} equations for two unknown functions $U(\xi)$, $N(\xi)$, namely,
\begin{gather*}
kn N+n\xi N' +(UN)'=0 ,\\
\left(mU+N'+ \frac{2m}{nk}(n \xi U' + UU')\right) \tau_0 \tau^{k+1}=0, \\
(n \xi U' + UU') a_0 \tau =0 .
\end{gather*}
This admits nontrivial solutions only if $k=1$,
\begin{gather*}
U= -n \xi, \qquad N= \frac{nm}{2} \big(\xi^2+N_0\big) ,
\end{gather*}
with the amplitudes and the time scaling factor becoming
\begin{gather*}
B = \tau_0 \tau ,\qquad A^2={\frac{2m\tau_0}{n}\tau+a_0}, \qquad F(\tau)^2= {{2mn}\tau_0\tau^5+a_0\tau^4} .
\end{gather*}
In this way, via the consistent ansatz~(\ref{tau-2}) we recover solutions of the form~(\ref{ovos}) after substituting the definitions of the independent variables $\xi$ and $\tau$ back into~(\ref{gss}). The more general form of these solutions, that is, solutions of the form~(\ref{ovo}), can be thought of as being obtained by means of a Galilean transformation applied to~(\ref{ovos}).

\subsection{The Lie symmetry}\label{subsection:Lie symmetry}
It is well known (see, e.g., \cite{BZ}) that self-similar solutions of a given PDE are often associated with its symmetries.
This is indeed the case for the solutions of the form (\ref{ovo}), as we presently show.

The Airy system~(\ref{SW1}) admits the Lie symmetry
\begin{gather}
 \eta_s=2 \eta-{x\eta_x},\qquad u_s=u-xu_x ,\label{Liesymm}
\end{gather}
i.e., $\eta_{ts}=\eta_{st}$ and $u_{ts}=u_{st}$, as can be shown by a straightforward computation.
The solutions of equations (\ref{Liesymm}) are
\begin{gather*}
\eta(x,s)={\rm e}^{2s} N(\xi) , \qquad u(x,s)={\rm e}^s U(\xi) ,
\end{gather*}
for generic functions $N$ and $U$ of the variable $\xi={x}/{{\rm e}^s} $. If $(\eta(x,t),u(x,t))$ is a solution of (\ref{SW1}), then for all $s\in{\mathbb R}$ we have that
\[
\eta(x,t;s)={\rm e}^{2s} \eta\big(x/{\rm e}^s,t\big) , \qquad u(x,t;s)={\rm e}^s u\big(x/{\rm e}^s,t\big)
\]
is also a solution of (\ref{SW1}). Let us expand the Airy fields as
\begin{gather}\label{uetatex}
\eta(x,t)=\gamma(t)x^2+\omega(t)x+\zeta(t)+R_\eta(x,t), \qquad u(x,t)=\alpha(t)x+\beta(t)+R_u(x,t) ,
\end{gather}
with $R_\eta(x,t)/x^2\to 0$ and $R_u(x,t)/x\to 0$ for $x\to 0$ with $t$ bounded.
Along the flow of the symmetry (\ref{Liesymm}) we have, for the layer thickness variable $\eta$,
\begin{align*}
\eta(x,t;s)&={\rm e}^{2s}\big(\gamma(t){x^2} {\rm e}^{-2s}+ \omega(t) x {\rm e}^{-s}+\zeta(t)\big)+x^2 \frac{R_\eta(\xi,t)}{\xi^2}\\ &
\simeq \gamma(t){x^2}+\omega(t) x {\rm e}^s+{\rm e}^{2s}\zeta(t),\qquad\mbox{for $\xi \to 0$}.
\end{align*}
Similarly, for the velocity field $u$ along the flow of (\ref{Liesymm}), we have
\begin{gather*}
u(x,t,s)\simeq \alpha(t)x+{\rm e}^{s}\beta(t).
\end{gather*}
Notice that, for $x$ bounded and $t$ in any interval of continuity for the coefficients of the expansion~(\ref{uetatex}), the limit $s\to \infty$ corresponds to $\xi\to 0$. Hence, while not being generally invariant along the $s$-flow, regular solutions to the Airy-SWE approach the finite-dimensional manifold of parabolic/linear fields asymptotically for large $s$. Thus, the relevance of the self-similar solutions above is further highlighted by these symmetry arguments.

\subsection{The bi-Hamiltonian structure of Airy-SWE}\label{bi-sub}
The Airy-SWE (\ref{SW0}) are a pair of quasi-linear PDEs. As such, it is well known that they can be integrated via the hodograph method. Their integrability can also be viewed in the sense of Liouville as they inherit a bi-Hamiltonian formulation (see, e.g., \cite{MN,NutkuPavlov}) from their NLS dispersionless interpretation, with the pair of compatible local Poisson structures given by
\begin{gather*}
P_0=-\left(
\begin{matrix}
0 &\partial_x \\
\partial_x & 0
\end{matrix}
\right), \qquad P_1=-\frac12 \left(
 \begin{matrix}
 \eta \partial_x + \partial_x \eta & u \partial_x \\
 \partial_x u & 2 \partial_x
 \end{matrix}
 \right).
\end{gather*}
Note that the compatibility between $P_0$ and $P_1$ is guaranteed by the existence of the vector field
\begin{gather}\label{Zeq}
Z\equiv (\dot \eta=0, \dot u=2) ,
\end{gather}
deforming $P_1$ into $P_0$ via Lie derivative. By means of the bi-Hamiltonian recursion method \cite{CFO,NutkuPavlov}, or by the inverse scattering method \cite{Zak}
a one-parameter family of conserved densities can be encoded in generator $h(u,\eta;z)$, in particular in the limit $z\to \infty$,
\begin{gather}\label{Caz}
h(u,\eta;z)=-\sqrt{\frac{(u-z)^2}{4}-\eta}+\frac{z-u}{2}=\sum_{i=1}^\infty \frac{h_i(\eta,u)}{z^i} .
\end{gather}
(The first of these conserved density is the fluid's mass $\eta$.) These quantities are in mutual Poisson bi-involution, that is,
\begin{gather*}
\left\{\int_\RR h_\ell\D x ,\int_\RR h_m\D x \right\}_{P_0}=\left\{\int_\RR h_\ell\D x ,\int_\RR h_m\D x \right\}_{P_1}=0 .
\end{gather*}
Let us introduce the vector fields of the hierarchy associated with~(\ref{Caz}),
\begin{gather*}
X_k:=P_0 \D \left(\int_\RR h_{k+1}\D x\right)=P_1\D \left(\int_\RR h_{k}\D x\right) ,
\end{gather*}
and the vector field $T_0$ given by (\ref{Liesymm}). This does not commute with all the $X_k$'s; however, as shown in \cite{ZM}, it generates a family of symmetries of the hierarchy as follows. Acting on $T_0$ with the recursion operator
\begin{gather*}
N=P_1 P_0^{-1}=
\frac{1}{2}\left(
\begin{matrix}
u & 2 \eta +\eta_x \partial_x^{-1} \\
2 & u_x \partial_x^{-1} +u
\end{matrix}
\right)
\end{gather*}
to define iteratively $T_k \equiv N^k T_0$, yields
\begin{gather*}
[T_i,T_j]=-(i-j) T_{i+j} \qquad \mbox{for all $i,j \ge 0$} ,
\end{gather*}
while, still thanks to the general properties of recursion operators,
\begin{gather*}
[X_i,T_j]=-j X_{i+j} .
\end{gather*}
Note that the hierarchy of $\tau$-symmetries generated (see \cite{FL,OFZR}) via the action of the recursion operator $N$ with $T_0$ as a seed does not coincide with the hierarchy of symmetries stemming from the vector field $Z$ defined by~(\ref{Zeq}).

\section{The reduced equations of motion}\label{link}
A natural generalization of the three-field self-similar solutions~(\ref{ovos}) is given by polynomial solutions of the form \cite{Ovs}
\begin{gather}\label{ovo2}
\eta(x,t)=\gamma(t) x^2+\omega(t) x +\zeta(t) , \qquad u(x,t)=\alpha(t) x+\beta(t) .
\end{gather}
The Airy system reduces to the following closed system of ODEs for the coefficients of the parabolic solution:
\begin{gather}
 \dot{\alpha}+\alpha^2+2\gamma=0 , \qquad \dot{\gamma}+3\alpha \gamma=0 , \qquad\dot{\zeta}+\alpha \zeta + \beta \omega=0,\nonumber\\
 \dot{\omega}+2\beta \gamma +2 \alpha \omega=0 , \qquad \dot{\beta}+\alpha \beta+ \omega=0 .\label{XVF}
\end{gather}

This system admits a symmetry (induced by the $x$-translational symmetry of the parent system) given by
 \begin{gather}\label{YVF}
 \alpha'=0 , \qquad
 \gamma'=0 , \qquad
 \zeta'=\omega , \qquad
 \omega'=2 \gamma , \qquad
 \beta'= \alpha .
 \end{gather}
As discussed in Section \ref{bi-sub}, the Airy system admits the family of mutually commuting conserved densities generated by (\ref{Caz}),
obtained by expanding $h(u,\eta;z)$ in powers of $z$ in the limit \smash{$z\to +\infty$}:
\begin{gather}
h(u,\eta;z)=\frac{\eta }{z}+\frac{\eta {u}}{z^2}
+\frac{\eta ^2+\eta {u}^2}{z^3}+\frac{3 \eta ^2 {u}+\eta {u}^3}{z^4}+\frac{2 \eta ^3+6 \eta ^2{u}^2+\eta {u}^4}{z^5}+\cdots. \label{devh}
\end{gather}
We can equip system (\ref{XVF}) with a sequence of conserved quantities from the Casimir~(\ref{devh}) of the Poisson pencil, by using the following construction, which is inspired by our previous works \cite{CFOPP, CFOPT}. The integration of the densities $h_i$ over the whole real line would yield ill-defined quantities for our field configurations (\ref{ovo2}). A physically sensible way to regularize them is to bound the range of $\eta$ by piecewise links to background constant states~(see, e.g.,~\cite{CFOPT}). Here, we choose the background constant to be zero, so that the fluid domain is bounded by the floor $\eta=0$ for $\gamma<0$ (and $\omega^2 -4 \gamma \zeta>0$). Thus, the support of the fluid is the finite interval
 \begin{gather}
 \label{xpm}
 x_- < x <x_+ , \qquad x_\pm = \frac{-\omega \mp \sqrt{\omega^2 -4 \gamma \zeta}}{2 \gamma} .
 \end{gather}
Notice that this construction is well defined since the equation for $\dot{\gamma}$ in (\ref{XVF}) implies that $\gamma$ cannot change sign during the time evolution.
 Integrating the quantities $h_i(\eta,u)$ between the two limits~(\ref{xpm}) yields a well defined sequence of conserved quantities. (These quantities can be extended to hold for different signs of the reduction parameters, which can then be interpreted physically by suitable normalizations, see, e.g., \cite{CFOPP}).
 From the density $\eta$ and the linear momentum density $\eta u$ one can construct the quantities
 \begin{gather*}
 H_1\equiv\int_{x_-}^{x_+} \eta \D x =\int_{x_-}^{x_+} \big(\gamma x^2+\omega x +\zeta\big) \D x = \frac{\big(\omega ^2-4 \gamma \zeta \big)^{3/2}}{6 \gamma^2} ,
 \end{gather*}
and
 \begin{gather*}
 H_2 \equiv \int_{x_-}^{x_+} \eta u \D x = \frac{\big(\omega ^2-4 \gamma \zeta \big)^{3/2} (2 \beta \gamma-\alpha \omega )}{12 \gamma ^3} ,
\end{gather*}
respectively. Similarly, the ``energy density'' $\eta u^2+\eta^2$ yields
 \begin{gather*}
 H_3 = {\frac {\big(\omega^2 -4\gamma\zeta \big)^{3/2} \big( 2{\alpha}^{2}
\gamma\zeta-3{\alpha}^{2}{\omega}^{2}+10\alpha\beta\gamma\omega-10{\beta}^{2}{\gamma}^{2}-8{\gamma}^{2}\zeta+2\gamma{\omega}^{2} \big) }{{60 \gamma}^{4}}},
\end{gather*}
while the fourth member of the sequence provides the conserved quantity
\begin{gather*}
H_4= - \big(\omega^2 -4\gamma\zeta \big)^{3/2} {\frac {( \alpha\omega-2
\beta\gamma ) \big( 3{\alpha}^{2}\gamma\zeta-2{
\alpha}^{2}{\omega}^{2}+5\alpha\beta\gamma\omega-5{\beta}^{2
}{\gamma}^{2}-12{\gamma}^{2}\zeta+3\gamma{\omega}^{2} \big) }{{60\gamma}^{5}}} ,
\end{gather*}
and so on to higher orders. By defining
\begin{gather*}
K_0\equiv (6 H_1)^{2/3}={\frac {{\omega}^{2}-4\gamma\zeta}{{\gamma}^{4/3}}} ,
\end{gather*}
a closer look at these first four quantities shows that they can be represented as
\begin{gather}
H_1=\dsl{}\frac16 K_0^{3/2},\qquad H_2=\dsl{\frac{1}{6} K_0^{3/2} \left(\beta -\frac{\alpha \omega }{2 \gamma }\right)} ,\nonumber\\
H_3=\dsl{K_0^{3/2}\left(\frac{1}{6}
\left(\beta -\frac{\alpha \omega }{2 \gamma }\right)^2+\frac{K_0}{120} \left(\frac{\alpha^2}{\gamma^{2/3}}-4\gamma^{1/3}\right)\right)} ,\nonumber\\
H_4=\dsl{K_0^{3/2} \left(\beta -\frac{\alpha \omega }{2 \gamma } \right) \left(\frac16 \left(\beta -\frac{\alpha \omega }{2 \gamma }\right)^2+\frac{K_0 }{40} \left(\frac{\alpha^2}{\gamma^{2/3}}-4\gamma^{1/3}\right)\right)} ,\label{Hamfindim}
\end{gather}
and, continuing on,
\begin{gather*}
H_5=K_0^{3/2} \left(
\frac16\left(\beta -\frac{\alpha \omega }{2 \gamma }\right)^{4}+
\frac{K_{{0}}}{20} \left(\beta -\frac{\alpha \omega }{2 \gamma }\right)^{2}\left(\frac{\alpha^2}{\gamma^{2/3}}-4\gamma^{1/3}\right)+ {\frac {{K_0}^2}{1120}}
 \left(\frac{\alpha^2}{\gamma^{2/3}}-4\gamma^{1/3}\right)^{2}\right)
\end{gather*}
etc. Since under the action of the canonical Poisson tensor $P_0$, the densities
\[
h_2 = \eta u \qquad\mbox{and}\qquad
\frac{h_3}2=\frac12 \eta\big(\eta^2+u^2\big)
\]
generate, respectively, $x$-translation and time evolution, the physical meaning of the quantities naturally appearing in (\ref{Hamfindim}) is clear:
\begin{itemize}\itemsep=0pt\item
$H_1\equiv \dsl{\frac16 K_0^{3/2}}$ is the total mass of fluid,
\item $\dsl{\beta -\frac{\alpha \omega }{2 \gamma }}$ is the center of mass speed, \item
$\dsl{\frac{{K_0}^{5/2}}{240}\left(\frac{\alpha^2}{\gamma^{2/3}}-4\gamma^{1/3}\right)}$ represents the energy of the motion in the center of mass frame.
\end{itemize}
 The structure of the so-obtained constants of the motion clearly suggests the opportunity of considering, as generators of the ring of invariant functions obtained by the Casimir of the Airy-SWE Poisson pencil, the simpler quantities
\begin{gather}
 K_0 =\frac{\omega ^2-4 \gamma \zeta }{\gamma^{4/3}},\qquad K_1 =\beta -\frac{\alpha \omega }{2 \gamma },\qquad K_2 =\frac{\alpha^2}{\gamma^{2/3}}-4\gamma^{1/3} . \label{ccon}
 \end{gather}
 These are defined on the manifold $M_5=\big\{(\alpha,\gamma,\zeta,\omega,\beta)\in{\mathbb R}^5\,|\, \gamma\ne 0\big\}$, without assumptions on the signs of $\gamma$ and $\omega^2 -4 \gamma \zeta$.

\subsection{Reductions for general symmetric smooth initial data}
When initial data for system~(\ref{SW0}) are analytic in $x$, local solutions as $|x|\to 0$ can be sought in the form of powers series in~$x$. Focussing for simplicity on the symmetric case $\eta(-x,\cdot)=\eta(x,\cdot)$, $u(-x,\cdot)=-u(x,\cdot)$, these are
\begin{gather}\label{Ssolnear0}
\eta(x,t)=\sum_{n=0}^\infty \eta_n(t)x^{2n} , \qquad u(x,t)=\sum_{n=0}^\infty u_n(t)x^{2n+1} ,
\end{gather}
where, as well known, the Cauchy--Kovalevskaya theorem (see, e.g.,~\cite{Evans}) assures that these power series have a finite radius of convergence, so that all coefficients are analytic in a neighbourhood of $t=0$. The exact solutions~(\ref{ovo2}) restricted to the invariant manifold $\omega=\beta=0$ are of course in this class, and are particular cases in that they derive from a hierarchy that truncates at the leading orders, $n=0,1$, so that $\eta$ and $u$ are second and first degree polynomials, respectively. Of course, this property is special, and the series~(\ref{Ssolnear0}) cannot in general be expected to truncate to a polynomial form; once substituted in~(\ref{SW0}) the series generate a recursive infinite hierarchy of ODEs for the coefficients $\eta_n(t)$ and $u_n(t)$. The first equations in the hierarchy are (cf.~(\ref{coeffODEs}))
\begin{gather}\label{paraeq}
\dot{u}_0+u_0^2+2\eta_1=0,\qquad \dot{\eta}_0+u_0\eta_0=0, \qquad \dot{\eta_1}+3(\eta_1 u_0+\eta_0 u_1)=0.
\end{gather}
Note that, when applied to zero velocity smooth initial data with a dry point~$\eta_0(0)=0$, then $\eta_0(t)=0$ for all times and these equations reduce to a closed system for $u_0$ and $\eta_1$ which can be solved exactly (see below). In fact, the presence of the dry point {\it linearizes} the evolution equations of all the pairs $\eta_{n+1}$ and $u_n$ with $n \geq 1$, whose coefficients and inhomogeneous terms depend solely on the ($\eta_{k+1},u_k$) pairs, $k<n$. Note that this linearization property induced by a dry point still requires solving an infinite sequence of ordinary differential equations, up to a~possible singularity time $t=t_s$, to determine the solution of the original PDE. A more direct analog of the self-similar solutions that truncate to second order polynomial also exists in the power series approach, which we examine next.

\subsection{More general algebraic reductions}
Power series solutions of the Airy system~(\ref{SW0}) admit a natural reduction that can be viewed as a generalization of the simple parabolic truncation~(\ref{ovo2}). Focussing once again for simplicity on the symmetric case, expressed by definitions~(\ref{Ssolnear0}) above, one can choose the primary constraint of either $u_n$ or $\eta_{n+1}$, $n>1$, to be identically zero (or, equivalently, just constant). With such a~constraint, the equations that determine the time dependence of the higher order coefficients~$u_l$ and~$\eta_{l+1}$ for all $l>n$ reduce to algebraic relations to the preceding set of coefficients with $l \leq n$. For instance, the case $n=1$ and $u_1=0$ yields system~(\ref{paraeq}) above, while with $n=2$ and setting $u_2\equiv 0$ yields the reduction
\begin{gather*}
\eta_3=-{1\over 2}u_1^2,\qquad u_3= {3\over {14\eta_0}}\big(5 u_1^2 u_0 -2\eta_2 u_1\big) ,
\end{gather*}
and so on. Thus, provided the series converge, the resulting solutions can be seen as finite dimensional reductions of the PDE even though they are not of polynomial type. The general non-symmetric case can be handled similarly with the appropriate choice of primary constraints. The geometric interpretation of these reductions and their study in the context of Hamiltonian structures goes beyond the aims of the present work and will be reported elsewhere.

\section{Explicit solutions and their physical meaning}\label{sec:explicit-solutions}

In this section we will solve explicitly the system (\ref {XVF}) {\it via} its conserved quantities, extending in this way a result obtained in \cite{CFOPT}. The explicit solution of the system is better obtained using a~set of physically relevant coordinates. These are given by the velocity gradient $\alpha$, the parabola curvature $\gamma$, the abscissa and the height of the vertex, respectively,
\[
\xi= -\frac{\omega}{2 \gamma},\qquad\mu=\zeta-\frac{\omega^2}{4 \gamma}
\]
(which can be used as alternatives to $\omega$ and $\zeta$), together with $\delta=\beta -\frac{\alpha \omega }{2\gamma}$. They are readily seen to be rectifying coordinates for the momentum vector field $Y$, in the sense that
\[
Y(\xi)={\xi}'=-1
\]
and the derivatives along $Y$ of the other coordinates vanish. We obtain the following representation for the vector field $X$:
\begin{gather*}
\dot{\alpha}+\alpha^2+2\gamma=0 ,\qquad \dot{\gamma}+3\alpha \gamma=0 ,\qquad \dot{\mu}+\alpha \mu=0 ,\qquad \dot{\xi}=\delta ,\qquad \dot{\delta}=0 .
\end{gather*}
The last two equations yield
\begin{gather*}
\delta(t)= \beta_0 -\frac{\alpha_0 \omega_0 }{2 \gamma_0 } \equiv \delta_0 ,\qquad
\xi(t)= \left( \beta_0 -\frac{\alpha_0 \omega_0 }{2 \gamma_0 } \right) t + \left( -\frac{\omega_0}{2 \gamma_0} \right) \equiv \delta_0 t + \xi_0 ,
\end{gather*}
that is, the motion of the abscissa of the parabola vertex is uniform. The equations for $\mu$ and $\gamma$ imply that these quantities maintain their initial sign. In particular, as expected by the physical meaning of $\eta$, if $\mu$ is initially positive, then it remains positive for all times (for as long as singularities do not develop).

The conserved quantity $K_2$ yields the relation
\begin{gather*}
\alpha^2 = 4 \gamma + K_2 \gamma^{2/3} ,
\end{gather*}
which coincides with that of the three-field reduction (\ref{ovos}). Then one obtains the solution (see \cite{CFOPP, CFOPT} for the case $\alpha_0=0$)
\begin{gather}
\alpha^2= 4 \gamma_0 \tau^3 + \big(\alpha_0^2 -4 \gamma_0\big) \tau^2 ,\qquad \gamma=\gamma_0 \tau^3 , \qquad \zeta=\zeta_0 \tau ,\label{gensol3cpi-bis}
\end{gather}
where $\tau(t)$ solves
\begin{gather*}
\dot{\tau}^2= 4 \gamma_0 \tau^5 + \big(\alpha_0^2 -4 \gamma_0\big) \tau^4 , \qquad \tau(0)=1,
\end{gather*}
and thus is related to $t$ by
\begin{gather}\label{tau-evolution}
t(\tau)=\frac{|\alpha _0| \tau -\sqrt{\alpha _0^2+4 \gamma _0 (\tau -1)}}{\tau \big(\alpha _0^2-4 \gamma _0\big)}+\frac{4 \gamma _0
\left(\mathrm{atanh}\sqrt{\frac{\alpha _0^2+4 \gamma _0 (\tau -1)}{\alpha _0^2-4 \gamma _0}}
-\mathrm{atanh} \left(\frac{|\alpha _0| }{\sqrt{\alpha _0^2-4\gamma _0}}\right)\right)}{\big(\alpha _0^2-4 \gamma _0\big){}^{3/2}}
\end{gather}
if the condition $\alpha _0^2 \neq 4 \gamma _0$ is fulfilled.

We close this section discussing the case $\alpha_0=\beta_0=0$ (i.e., the fluid is initially at rest) and $\gamma_0>0$.
From (\ref{tau-evolution}) it can be checked that $\tau\to +\infty$ if
\[
t\to t_s\equiv\frac\pi{4\sqrt{\gamma_0}}.
\]
Hence (\ref{gensol3cpi-bis}) shows that there is a blow-up of $\alpha$, $\gamma$, and $\zeta$ at the finite time $t_s$.
Therefore we can see the variable change $\tau=\tau(t)$ as an analogue of the classical regularizing
Kustaanheimo-Stiefel transformation for the Kepler problem (see, e.g., \cite{Cor} and the references quoted therein).

 \section{The bi-Hamiltonian structure} \label{sezbih}
 In this section we construct a bi-Hamiltonian structure for our system, that is, a pair of compatible Poisson tensors $P$ and $Q$ on $M_5$ fulfilling the Lenard--Magri
\cite{GGKM74, Mag78}, or Gelfand--Zakharevich~\cite{GZ93} relations\footnote{See, e.g., \cite{MCFP04} and references quoted therein for a gentle introduction to this theory, and \cite{PS05} for the historical origin of the term {\em Lenard--Magri} relations/sequences/chains.}
 \begin{gather}\label{ladder}\begin{split}& \boxed{
\xymatrix{
 & \D K_0 \ar[dl]_P \ar[dr]^Q & & \D K_1 \ar[dl]_P \ar[dr]^Q & & \D K_2 \ar[dl]_P \ar[dr]^Q & \\
 0 & & Y & &X & & 0
}}\end{split}
\end{gather}
where, here and below, $X$ and $Y$ are the vector fields (\ref{XVF}) and (\ref{YVF}). In other words, we require that $P$ satisfies
 \begin{gather}\label{nonhP}
P \D K_0=0 , \qquad
P \D K_1=Y , \qquad
P \D K_2=X ,
\end{gather}
while $Q$ satisfies
 \begin{gather}\label{nonhQ}
Q \D K_0=Y , \qquad
Q \D K_1=X , \qquad
Q \D K_2=0 ,
\end{gather}
where the $K_j$'s are defined in (\ref{ccon}).

A consequence of (\ref{ladder}) being fulfilled is that the three Hamiltonians are in mutual involution with respect to both brackets, that is,
\begin{gather}\label{homoPQ}
\{K_i, K_j\}_P=\langle \D K_i, P \D K_j\rangle=0,\qquad \{K_i, K_j\}_Q=\langle \D K_i, Q \D K_j\rangle=0,\qquad i,j=0,1,2.\!\!
\end{gather}
To show this, it is convenient to introduce another set of coordinates. In fact, $K_0$ and $K_1$ are, respectively, linear in $\zeta$ and $\beta$, and suggest for coordinates in $M_5$ the set
 \begin{gather} \label{nvar}
 (\alpha, \sigma, \kappa, \omega, \delta) ,
 \end{gather}
 where
\begin{gather}\label{nvar2}
\sigma={\gamma^{1/3}} , \qquad \kappa= -\frac{\omega ^2-4 \gamma \zeta }{4 \gamma^{4/3}} , \qquad \delta= \beta -\frac{\alpha \omega }{2 \gamma } ,
\end{gather}
with inverse relations
\begin{gather*}
\gamma=\sigma^3,\qquad \beta=\delta+{\frac {\alpha\omega}{{2 \sigma}^{3}}}, \qquad \zeta=\sigma\kappa+{\frac {{\omega}^{2}}{{4 \sigma}^{3}}} .
\end{gather*}
The conserved quantities now read
 \begin{gather*}
 K_0 =\kappa,\qquad K_1 =\delta,\qquad K_2 =\frac{\alpha ^2 }{\sigma^2} -4 \sigma, 
 \end{gather*}
and system~(\ref{XVF}) becomes
\begin{gather}
\dot{\alpha}+\alpha^2+2 \sigma^3 =0 , \qquad \dot{\sigma}+\alpha \sigma=0 , \qquad \dot{\kappa}=0 , \qquad
\dot{\omega}+3 \alpha \omega+2 \delta \sigma^3 =0 , \qquad \dot{\delta}=0 .\label{5ODEsigma}
\end{gather}
Moreover, the symmetry (\ref{YVF}) is
 \begin{gather*}
 \alpha'=0 , \qquad \sigma'=0 , \qquad \kappa'=0 , \qquad \omega'=2 \sigma^3 , \qquad \delta'= 0 \label{momconsymmsig}
 \end{gather*}
(which shows that $K_j'=0$ for $j=1,2,3$, so that $X$ and $Y$ share these three constants of the motion).

To find the descending Poisson structure $P$ in~(\ref{ladder}), we first consider the bivector
\begin{gather*}
P':=Y\wedge\partial_\delta+X\wedge X_\alpha ,
\end{gather*}
where $X_\alpha=\dsl{{\sigma^2}/({2 \alpha})}\partial_\alpha.$ Since $X_\alpha(K_2)=\partial_\delta(\delta)=1$ and
$\partial_\delta(K_2)= X_\alpha(\delta)=0$, relations~(\ref{nonhP}) are indeed fulfilled.
Note that $P'$ is {\em not} a Poisson bivector. As well known, the Jacobi condition for a bivector $P$ to give rise to a Poisson structure is encoded by the {\em Schouten bracket} $[P,P]_S$ as
\begin{gather*}
\{f_1,\{f_2,f_3\}_P\}_P+\{f_2,\{f_3,f_1\}_P\}_P+\{f_3,\{f_1,f_2\}_P\}_P=-\frac12 \langle \D f_1\wedge \D f_2\wedge \D f_3, [P,P]_S\rangle.
\end{gather*}
Using standard formulas of Poisson calculus (see, e.g., \cite{Izu}) that express the Schouten bracket of decomposable bivectors in term of ordinary Lie brackets of vector fields as
\begin{gather*}
[X_1\wedge X_2, Y_1\wedge Y_2]_S= -[X_1,Y_1]\wedge X_2\wedge Y_2+[X_1,Y_2]\wedge X_2\wedge Y_1 +[X_2,Y_1]\wedge X_1\wedge Y_2\\
\hphantom{[X_1\wedge X_2, Y_1\wedge Y_2]_S=}{} -[X_2,Y_2]\wedge X_1\wedge Y_1 ,
\end{gather*}
and taking into account that the only non vanishing Lie brackets among the four vector fields entering the definition of $P'$ are
\begin{gather*}
 [X,X_\alpha]=-\frac{\sigma^2}{2 \alpha^2} (X+\delta Y )\qquad\text{and}\qquad [X,\partial_\delta]=Y ,
\end{gather*}
we can easily compute
\begin{gather}\label{ScPpr}
[P',P']_S=[X\wedge X_\alpha, X\wedge X_\alpha]_S=2[X,X_\alpha]\wedge X_\alpha\wedge X
=- \frac{\sigma^2 \delta }{\alpha^2} Y\wedge X_\alpha\wedge X.
\end{gather}
However, note that the bivector $R$ defined by
\begin{gather*}
R\equiv Y\wedge X
\end{gather*}
has a kernel spanned by $(\D K_0, \D K_1, \D K_2)$, since the $K_j$'s are conserved quantities for both $Y$ and $X$. Thus, any bivector of the form
\begin{gather}\label{Pf}
P_f:=P'+f R,
\end{gather}
where $f$ is an arbitrary function on $M_5$, still satisfies the conditions (\ref{nonhP}). Hence, we can use this ``extra'' degree of freedom to seek a {\em Poisson\/} structure among the family of bivectors (\ref{Pf}). Indeed, another straightforward computation shows that
\begin{gather*}
[P_f,P_f]_S=-2Y(f) Y\wedge \partial_\delta\wedge X+\left(2 X(f)-\frac{\sigma^2 \delta}{\alpha^2}\right)Y\wedge X_\alpha\wedge X .
\end{gather*}
Since $X$, $Y$, $X_\alpha$, $\partial_\delta$ together with $\partial_\kappa$ (which will be used for the second Poisson tensor $Q$ below) are a (local) basis for the sections of $T M_5$, we get that $f$ must satisfy the differential system
\begin{gather*}
\left\{
\begin{array}{@{}lcl}
Y(f)=0& \Longleftrightarrow &\sigma^3 f_\omega=0,\vspace{1mm}\\
X(f)=\dsl{\frac{\sigma^2 \delta}{2\alpha^2}}&\Longleftrightarrow &-\big(\alpha^2+2\sigma ^3\big)f_\alpha-\alpha\sigma f_\sigma-\big(3\alpha\omega+2\delta\sigma^3\big)f_\omega=\dsl{\frac{\sigma^2 \delta}{2\alpha^2}},
\end{array}
\right.
\end{gather*}
which turns out to be explicitly solvable. Indeed, this reduces to the linear equation
\begin{gather}\label{chareqp}
-\big(\alpha^2+2\sigma ^3\big)f_\alpha-\alpha\sigma f_\sigma=\dsl{\frac{\sigma^2 \delta}{2\alpha^2}},
\end{gather}
 for a function of three variables $f(\alpha,\sigma,\delta)$. This equation can be solved by the method of characteristics.
Using $\sigma$ as characteristic parameter, we have
\begin{gather*}
 \frac{\D f(\alpha(\sigma),\sigma,\delta(\sigma))}{\D \sigma} +\dsl{\frac{\sigma \delta}{2\alpha(\sigma)^3}}=0 , \qquad
 f(\alpha(\sigma_0),\sigma_0,\delta(\sigma_0))=f(\alpha_0,\sigma_0,\delta_0),
\end{gather*}
where the characteristics are the solutions of the Cauchy problems
\begin{gather*}
\frac{\D \alpha }{\D \sigma}= \frac{\alpha^2+2 \sigma^3}{\alpha \sigma} , \qquad
\frac{\D \delta}{\D \sigma}=0 , \qquad \alpha(\sigma_0)=\alpha_0 , \qquad \delta(\sigma_0)=\delta_0 .
\end{gather*}
The solutions of this ODEs system are
\begin{gather}
\frac{\alpha^2}{\sigma^2}=\frac{\alpha_0^2+4 \left(\sigma -\sigma_0\right) \sigma_0^2}{\sigma _0^2}, \qquad \delta=\delta_0,
\label{carf}
\end{gather}
and the variation of $f$ along the characteristics is given by
\begin{gather}\label{presolf}
 \frac{\D f(\alpha(\sigma),\sigma,\delta(\sigma))}{\D \sigma} + \dsl{\frac{\sigma \sigma_0^3 \delta}{2\sigma^3 \big(\alpha_0^2+4 (\sigma -\sigma_0 ) \sigma_0^2\big)^{3/2}}} =0.
\end{gather}
Using (\ref{carf}) and (\ref{presolf}) one can find that a possible solution of (\ref{chareqp}) is
\begin{gather*}
f(\alpha,\sigma,\delta)=\dsl{-6{\sigma}^{5}\delta
\frac{\mathrm{atanh}
\big( {\frac {\alpha}{\sqrt {{\alpha}^{2}-4{\sigma}^{3}}}} \big) }{\big( {-}4{\sigma}^{3}
+{\alpha}^{2} \big) ^{5/2}}+\frac12{\frac {{\sigma}^{2}\delta \big( 8
{\sigma}^{3}+{\alpha}^{2} \big) }{\alpha \big({\alpha}^{2} -4{\sigma}^{3}\big) ^{2}}}},
\end{gather*}
and the matrix structure of the tensor $P_f$ is given by
\begin{gather}
 \left( \begin{matrix}
 0&\frac12{\sigma}^{3}&0&P_f^{14}&0\vspace{1mm} \\
 -\frac12{\sigma}^{3}&0&0&-2f{\sigma}^{4}\alpha&0\vspace{1mm} \\
 0&0&0&0&0\vspace{1mm} \\
 -P_f^{14}&2f{\sigma}^{4}\alpha&0&0&2{\sigma}^{3}\vspace{1mm} \\
 0&0&0&-2{\sigma}^{3}&0\end{matrix} \right),
\label{finalP}
\end{gather}
with $P_f^{14}={{\frac {{\sigma}^{5}\delta}{\alpha}}}+\frac32{\sigma}^{2}\omega-2f \big( 2{\sigma}^{3}+{\alpha}^{2} \big) {\sigma}^{3}$.
By similar arguments, one can show that system~(\ref{5ODEsigma}) is Hamiltonian with respect to the second structure
\begin{gather}
\left(\begin{matrix}
 0 & 0 & 0 & \dsl{\frac{g \big(2 \sigma ^3+\alpha ^2\big)}{\alpha \sigma }} & -2 \sigma ^3-\alpha ^2 \vspace{1mm} \\
 0 & 0 & 0 & g & -\alpha \sigma \vspace{1mm} \\
 0 & 0 & 0 & -2 \sigma ^3 & 0 \vspace{1mm} \\
 \dsl{-\frac{g \big(2 \sigma ^3+\alpha ^2\big)}{\alpha \sigma } }& -g & 2 \sigma ^3 & 0 & -2 \delta \sigma ^3-3 \alpha \omega \vspace{1mm} \\
 2 \sigma ^3+\alpha ^2 & \alpha \sigma & 0 & 2 \delta \sigma ^3+3 \alpha \omega & 0 \vspace{1mm} \\
\end{matrix}\right),\label{Q-old}
\end{gather}
where $g=g(\alpha,\sigma)$ solves the equation
\begin{gather}\label{chareqq}
\big(\alpha^2+2\sigma ^3\big)g_\alpha+\alpha\sigma g_\sigma=1.
\end{gather}
We can rephrase the arguments leading to the tensor $P_f$ of (\ref{finalP}) for the second Poisson tensor~$Q$ as follows. A suitable bivector satisfying (\ref{nonhQ}) is given by
\begin{gather*}
Q':=Y\wedge \partial_\kappa+X\wedge\partial_\delta.
\end{gather*}
As before, $Q'$ is not a Poisson tensor, since $[Q',Q']_S=2 Y\wedge \partial_\delta\wedge X$. Defining
\begin{gather*}
Q_g:=Q'+g R
\end{gather*}
and imposing the vanishing of the Schouten bracket of $Q_g$ with itself (and taking into account that $\partial_\kappa$ has vanishing Lie bracket with the other vector fields appearing in $Q_g)$ yields the differential system
\begin{gather*}
\left\{
\begin{array}{@{}lcl}
Y(g)=0& \Longleftrightarrow &\sigma^3 g_\omega=0,\vspace{1mm}\\
X(g)=-1&\Longleftrightarrow &\big(\alpha^2+2\sigma ^3\big)g_\alpha+\alpha\sigma g_\sigma+\big(3\alpha\omega+2\delta\sigma^3\big)g_\omega=1,
\end{array}
\right.
\end{gather*}
which reduces to~(\ref{chareqq}). Its simplest solution is
\begin{gather*}
g(\alpha, \sigma)=4{\sigma}^{3}
\frac{\mathrm{atanh} \big( {\frac {\alpha}{\sqrt {{\alpha}^{2}-4{\sigma}^{3}}}} \big)}{\big( {\alpha}^{2}-4{\sigma}^{3} \big) ^{3/2}}
-{\frac {\alpha}{{\alpha}^{2}-4{\sigma}^{3}}}.
\end{gather*}
It turns out that requiring the compatibility of $P_f$ and $Q_g$ does not add any conditions on $f$ and $g$. In fact
\begin{gather*}
[P_f, Q_g]_S=\left(X(f)-\dsl{\frac{\delta\sigma^2}{2 \alpha^2}}\right) Y\wedge X\wedge\partial_\delta+\left(X(g)+1\right)Y\wedge X_\alpha\wedge X=0 ,
\end{gather*}
so that the tensors $P_f$ given in (\ref{finalP}) and $Q_g$ given in (\ref{Q-old}) endow $M_5$ with the structure of a~bi-Hamiltonian manifold.

Let us stress that, by construction, the Lenard--Magri chain (\ref{ladder}) follows and, consequently, the bi-involution relations~(\ref{homoPQ}). Actually, all the functions $H_j$, obtained via the integration process described in Section~{\ref{link}}, are in bi-involution, since they are constants of the motion for both $X$ and $Y$, and therefore
\begin{align*}
\{H_j, H_j\}_P& =\langle \D H_i, P \D H_j\rangle\\
& =Y(H_i) \partial_\delta(H_j)+X(H_i) X_\alpha(H_j)+f X(H_i)Y(H_j)-(i \leftrightarrow j)=0 .
\end{align*}
Note that this implies that $H_j$, $j\ge 3$, must be functionally dependent on $K_0$, $K_1$, and $K_2$, given that $P$ is rank four.

\section{Reductions}\label{redux}
In this final section we discuss how to further reduce the $5$-dimensional system of ODEs (\ref{XVF}) to physically significant submanifolds. The first case we consider is that of parity conserving solutions, obtained by ``killing'' the Galilean symmetry (\ref{YVF}). This is interesting since, by the arguments of Section~\ref{sec:explicit-solutions}, such a reduction provides the most relevant information on the evolution. The second case we consider next, that of the submanifold given by ``linear-linear'' initial data obtained by setting $\gamma=0$, is somewhat trickier. Although the integration of the resulting ODEs can be most easily performed in an iterative way, the Hamiltonian functions of the system (\ref{XVF}) blow up in this limit, and so an {\em ad hoc} procedure is needed to recover them.

\subsection{The reduction to the parity conserving solutions}
Parabolic-linear solutions of the form
\begin{gather}
 \eta(x,t)= \gamma(t) x^2+\zeta(t) , \qquad u(x,t)= \alpha(t) x , \label{numuga}
\end{gather}
of the Airy-SWE were first introduced in \cite{Ovs, Tal} and studied in \cite{BKK03, CFOPP,CFOPT}, where an explicit description of the solutions starting with zero velocities was given. Solutions (\ref{numuga}) are obviously a particular case, given by $\eta(-x,t)=\eta(x,t)$ and $u(-x,t)=-u(x,t)$, of the ones studied in the previous sections, and the three coefficients $(\alpha,\gamma,\zeta)$ evolve according to the following system of ODEs,
\begin{gather}
 \dot{\alpha}+\alpha^2 +2\gamma =0 , \qquad \dot{\gamma}+3 \alpha\gamma=0 , \qquad \dot{\zeta}+\alpha \zeta=0 .\label{coeffODEs}
\end{gather}
This is the reduction of the 5-field system discussed so far to the invariant manifold $M_3$ obtained by setting $\beta=\omega=0$ (that is,
$\delta=\omega=0$) in the original system.
The adapted coordinates and the coordinate change described in (\ref{nvar})--(\ref{nvar2}) now read
\[
\sigma=\gamma^{1/3},\qquad \kappa=\zeta\gamma^{-1/3},\qquad \text{with inverse}\qquad\gamma=\sigma^3,\qquad \zeta=\kappa\sigma.
\]
We now have two Hamiltonians, namely, $\kappa$ and $H\equiv K_2$. Note that the vector field $Y$ defined above is not tangent to $M_3$, as $x$-translations move the vertex away from the origin.
The time evolution (given by the vector field $X_3$, that is, $X$ restricted to $M_3$) reads
\begin{gather*}
\dot{\alpha}+\alpha^2+2 \sigma^3 =0, \qquad \dot{\sigma}+\alpha \sigma=0, \qquad \dot{\kappa}=0.
\end{gather*}
As far as the Poisson structures are concerned, the first one (fulfilling $P_3 \D \kappa=0, P_3 \D H= X_3$) is just the restriction of
the previously found $P_f$, that is
\[
P_3=X_3\wedge X_\alpha,
\]
where $X_\alpha\equiv \dsl{{\sigma^2}/({2\alpha})}\partial_\alpha$ is now seen as a vector field on $M_3$. Being the restriction of a Poisson structure, $P_3$ is automatically a Poisson structure (see also~(\ref{ScPpr})). Notice that $P_3$ can also be seen as the projection of $P_f$ along the distribution $D$ spanned by $Y$ and $\partial_\delta$.

The restriction of the second Poisson tensor $Q_g$ to $M_3$ does not exist. Its projection along $D$ is the vanishing bivector, so it is useless for our purposes.
However, a second Poisson tensor for the (shortened) Lenard--Magri chain
 \begin{gather*}\boxed{
\xymatrix{
 & \D \kappa \ar[dl]_{P_3} \ar[dr]^{Q_3} & & \D H \ar[dl]_{P_3} \ar[dr]^{Q_3} & \\
 0 & & X & &0}}
\end{gather*}
can be easily found to be
\[
Q_3=X_3\wedge \partial_\kappa.
\]
It can be checked that $P_3$ and $Q_3$ are compatible.

\subsection{The reduction to linear-linear solutions}
The submanifiold $\gamma=0$ is invariant for the ODEs~(\ref{XVF}).
The restriction of the vector field to this submanifold of linear-linear configurations of the form
\begin{gather*}
\eta(x,t)=\omega(t) x+\zeta(t), \qquad u(x,t)=\alpha(t) x+\beta(t)
\end{gather*}
is
\begin{gather}\label{X4}
\dot{\alpha}+{\alpha}^{2}=0,\qquad \dot{\zeta}+\alpha\zeta+\beta=0,
\qquad\dot{\omega}+2\alpha\omega=0,\qquad\dot{\beta}+\alpha\beta+\omega=0,
\end{gather}
and the Galilean vector field restricts to
\begin{gather}\label{Y4}
\alpha'=0 ,\qquad {\zeta}'=\omega ,\qquad \omega'=0 , \qquad {\beta}'=\alpha.
\end{gather}
System (\ref{X4}) can be integrated via a straightforward inductive process, starting from the first equation, substituting into the third, then into the fourth and finally into the second. The solution of the Cauchy problem is
\begin{alignat*}{3}
& \alpha (t) ={\frac {\alpha_{{0}}}{\alpha_{{0}} t+
1}},\qquad && \beta (t) =-{\frac {\omega_{{0}}\ln ( \alpha_{{0}} t+1) -\beta_{{0}}\alpha_{{0}}}{\alpha_{{0}}
( \alpha_{{0}} t+1) }},& \\
& \omega (t) ={\frac {\omega_{{0}}}{
 ( \alpha_{{0}} t+1) ^{2}}},\qquad && \zeta (t) =
 {\frac {\zeta_{{0}}}{\alpha_{{0}} t+1}}-{\frac {\omega_{{0}} (
\beta_{{0}}\alpha_{{0}}-\omega_{{0}}) t }{\alpha_{{0}} ( \alpha_{{0}} t+1) ^{2}}}-{\frac {\ln ( \alpha_{{0}} t+1
) {\omega_{{0}}}^{2}}{{\alpha_{{0}}}^{2} left( \alpha_{{0}} t+1) ^{2}}}.&
\end{alignat*}
The vector field $X_4$ given by (\ref{X4}) admits three global constants of the motion:
\begin{gather*}
H^{l}_1\equiv {\frac {{\alpha}^{2}}{\omega}}, \qquad H^l_2\equiv{\frac {\alpha\mu}{\omega}}-\beta+{\frac {\omega}{\alpha}},\qquad
H^l_3\equiv{\frac {\alpha\beta}{\omega}}-\ln |\alpha| .
\end{gather*}
The first two of them are shared with the Galilean vector field $Y_4$ defined by (\ref{Y4}). In complete analogy with the procedure discussed in Section \ref{sezbih}, one can find a pair of Poisson tensors generating the pair of vector fields (\ref{X4}), (\ref{Y4}) from $H^l_1$ and $H^l_2$. They are given by
\begin{gather*}
P_1=Y_4\wedge \frac{\partial}{\partial H^l_1}+X_4\wedge\frac{\partial}{\partial H^l_2}-\frac1{\alpha} Y_4\wedge X_4
\end{gather*}
and
\begin{gather*}
P_2=Y_4\wedge \frac{\partial}{\partial H^l_2}+X_4\wedge\frac{\partial}{\partial H^l_1}+\left({\frac {\mu}{{\alpha}^{2}}}+{\frac {\ln |\alpha| }{\big({{H^l_1}}\big)^{2}}}\right)Y_4\wedge X_4 ,
\end{gather*}
where we have used $\big(\alpha,\mu,H^l_1,H^l_2\big)$ as coordinates.

\section{Conclusions}
In this work, we have studied the geometric structure of self-similar solutions of the second kind of the Airy system~(\ref{SW0}). Of course, while the Airy system is used as an example, given its relevance from the physical viewpoint as it lies at the intersection of hydrodynamics, optics and condensed matter, our approach can be applied to more general setups that share the same fundamental structure. For instance, multilayer systems of different density fluids admit polynomial solutions whose proper geometric interpretation could be carried out similarly, as well as extensions to more spatial dimensions. Further generalizations naturally arise as well, extending the Hamiltonian degrees of freedom for the finite system of ODEs to numbers higher than two. As seen in Section~\ref{link}, the formalism developed herein could be extended beyond the case of polynomials of second degree to power series solutions. These and other topics are currently under investigation and will be reported in the future.

\subsection*{Acknowledgments}
RC and MP thank the {Dipartimento di Matematica e Applicazioni} of Universit\`a Milano-Bicocca for its hospitality. GF, GO, and MP thank the Carolina Center for Interdisciplinary Applied Mathematics at the University of North Carolina for hosting their visits in 2018. This work was supported by the National Science Foundation under grants RTG DMS-0943851, CMG ARC-1025523, DMS-1009750, DMS-1517879, the Office of Naval Research under grants N00014-18-1-2490 and DURIP N00014-12-1-0749. This project has also received fundings under grant H2020-MSCA-RISE-2017 Project No.~778010 IPaDEGAN. All authors gratefully acknowledge the auspices of the GNFM Section of INdAM under which part of this work was carried out. Finally, thanks are also due to the anonymous referees for useful comments and suggestions for further references (e.g.,~\cite{BKK03,Tal,Zak}). Their work improved the final form of the present paper.

\pdfbookmark[1]{References}{ref}
\LastPageEnding

\end{document}